\documentclass[aps,prb,groupedaddress,twocolumn,10pt]{revtex4-1}

\usepackage{color,soul}

\begin{document}

\newcommand{\be}{\begin{equation}}
\newcommand{\ee}{\end{equation}}
\newcommand{\bea}{\begin{eqnarray}}
\newcommand{\eea}{\end{eqnarray}}

\title{Viable phenomenologies of the normal state of cuprates}

\author{D. V. Khveshchenko}

\affiliation{Department of Physics and Astronomy, University of North Carolina, Chapel Hill, NC 27599}

\begin{abstract}
We revisit the problem of constructing an elusive 
scaling theory of the strange metal phase of the cuprates. 
By using the four robust experimentally 
established temperature dependencies as the constitutive relations
we then predict the scaling behaviors of a 
number of other observables, all those for which the reliable data are available
being in agreement with experiment. 
Such predictions are also contrasted against the recent proposal 
inspired by the holographic approach, thus 
allowing one to critically assess the status of the latter. 
\end{abstract}

\maketitle

The mystery of the normal state of the cuprate superconductors has long remained a challenge 
defying numerous attempts of its theoretical understanding. 
In the continuing absence of a satisfactory microscopic description, a more modest goal 
would be that of constructing a phenomenological description capable of accounting 
for the observed anomalous transport properties. 

Recently, there has been an upsurge of interest in the transport properties of the cuprates and other 'strange metals' inspired by a proliferation of
 ideas pertaining to the novel interdisciplinary field of holography \cite{ads}. 
While relying on a bold assumption of its potential (albeit, so far, unproven) 
applicability well outside the native realm of 
the string/supergravity/gauge theories, the holographic approach 
prompted a renewed quest into the general transport properties 
of strongly correlated systems. 

In particular, such analyses are seeking to establish some 
general relations between and universal bounds for the transport coefficients 
which could remain valid, regardless of the (in)applicability of the generalized holographic conjecture itself.
Specifically, such relations are expected to hold in the hydrodynamic regime
dominated by inelastic scattering between the constituent fermions. 
 
Envisioned among the candidate systems, are not only the usual suspects such as quark-gluon plasma and cold Fermi gas near unitarity but also, conceivably, such documented strongly correlated condensed matter systems as the cuprates, heavy fermion materials, graphene, etc.  
Adding to the intrigue, the recent ARPES data on the compound 
$BiSrCaCuO$ were argued to demonstrate 
an unusually low value of the shear viscosity-to-entropy ratio which is comparable to those found in the quark-gluon plasma and unitary Fermi gas \cite{bnl}.

The early phenomenologies of the cuprates focused on the seemingly irreconcilable dichotomy between the robust power-law behaviors of the longitudinal conductivity
\be
\sigma\sim T^{-1}
\ee
and the (electrical) Hall angle
\be
\tan\theta_H\sim T^{-2}
\ee
In the popular scenaria, Eqs.(1,2) were argued to imply the existence of two distinct 
scattering times: $\tau_{tr}\sim 1/T$ and  $\tau_{H}\sim 1/T^2$ which characterize the relaxation of either 
longitudinal vs transverse \cite{anderson}
or charge-symmetric vs anti-symmetric \cite{coleman} currents.
Yet another insightful proposal was put forward in the framework of the marginal Fermi liquid phenomenology \cite{varma}. 

One more piece of the puzzle is provided by the magnetoresistivity
which violates the conventional Kohler's law 
$\Delta\rho/\rho\sim \rho^2$, while still obeying 
its replacement \cite{kohler}
\be
{\Delta\rho\over \rho}\sim \theta^2_H\sim T^{-4}
\ee
The above anomalous transport properties have to be contrasted 
against the rather conventional thermodynamic ones, including the Fermi-liquid-like specific heat 
(except for a possible logarithmic enhancement \cite{specificheat})
\be
C\sim T
\ee
The simple power-law temperature dependencies (1-4) suggest the possibility   
of an extended one-parameter scaling regime. Physically, such a behavior would generally be expected 
to occur in the vicinity of some incipient quantum critical point
characterized by the absence of a competing energy scale, other than temperature.
In the context of the cuprates, a number of the potentially
viable quantum critical transitions have been discussed, their list including superconducting, spin, charge, nematic, as well as other, even more exotic, instabilities. 

Under the above assumption, the exponents in Eqs.(1-4) can be readily related to
the dimensions of the underlying fields and currents.  
The latter can deviate from their canonical ('engineering') values because 
of a potentially non-trivial dynamical exponent $z$, as well as 
a scale-dependent ('running') charge $e$ with a 
dimension $\Delta_e$ which takes into 
account the renormalization effects  
('vertex corrections') \cite{dvk,karch}.

The dimensions of the scalar and vector electromagnetic potentials and the corresponding electric and magnetic fields read 
\be
[A_0] = z-\Delta_e,~~[{\bf A}]=1-\Delta_e,
\ee
$$
[{\bf E}] = z+1-\Delta_e,~~
[{\bf B}]=2-\Delta_e
$$
It is worth emphasizing here that a clear distinction has to be made between $A_0$ and the chemical potential $\mu$ with the dimension $[\mu]=[\epsilon]=[T]=z$ \cite{dvk} (in contrast, 
identifying the two, as in Refs.\cite{karch,hk}, 
makes it rather problematic, among other things, to even set up       
such dimensionless combinations as $(\epsilon-\mu)/T$). 

In turn, the dimensions of the electrical (${\bf J}$) and thermal 
(${\bf Q}$) currents carried by the species of charge $e$, energy $\epsilon$,
velocity $v$, and (number) density $n$ are given by the relations 
\be
[{\bf J}]=[env]=\Delta_n+\Delta_e+z-1,
\ee
$$
[{\bf Q}]=[\epsilon nv]=\Delta_n+2z-1
$$
The linear thermoelectric response 
is then described in terms of a trio of 
the fundamental kinetic coefficients
\be
{\bf J}={\hat \sigma} {\bf E}-{\hat \alpha} {\bf \nabla}T
\ee
$$
{\bf Q}=T{\hat \alpha} {\bf E}-{\hat {\kappa}} {\bf \nabla}T
$$
where the Onsager's symmetry is taken into account
and the off-diagonal entries of the $2\times 2$ matrices 
represent the Hall (off-diagonal) components of the corresponding conductivities. 

Setting up the scaling relations for reproducing Eqs.(1-4) and other experimentally observed algebraic dependencies one has to properly account for the implications of the time reversal and particle-hole symmetries.
A compliance with such symmetries dictates the 
following dependencies of the Hall angle, magnetoresistivity, 
conductivity, and specific heat on the temperature, 
magnetic field, and chemical potential
\be
{\sigma}\sim T^{(\Delta_n+2\Delta_e-2)/z},
\ee
$$
\tan\theta_H\sim B^{\beta}\mu^{\lambda}e^{\gamma+\delta} 
T^{-(2\beta+z\lambda+(-\beta+\gamma+\delta)\Delta_e)/z}, 
$$
$$
{{\Delta\rho}\over \rho}
\sim 
B^{2\beta}e^{2\delta} 
T^{-2(2\beta+(-\beta+\delta)\Delta_e)/z},
$$
$$
C\sim T^{\Delta_n/z}
$$
Here an (in general, non-analytical) dependence on $B$ and $\mu$ is included to account for
the breaking of time-reversal and particle-hole symmetries, respectively, 
the corresponding exponents being $\beta$ and $\lambda$.
Furthermore, adding such factors also requires one to introduce (in general, 
unrelated) powers $\gamma$ and $\delta$ of the effective charge 
which, given the possibility of a non-zero dimension $\Delta_e$, 
may be needed for a proper power-counting.

After having matched Eqs.(8) with (1-4)
and solved for $z,\Delta_e, \Delta_n$ one 
can also check for agreement with the other measured observables, such as  
thermopower $S=\alpha/\sigma$, Hall Lorentz number $L_H=\kappa_H/T\sigma_H$,
and Nernst coefficient $\nu_N=(\alpha_H\sigma-\alpha\sigma_H)/B(\sigma^2+\sigma^2_H)$ 
which scale as 
\be
S\sim \mu^{-1}T^{(z-\Delta_e)/z},
\ee
$$
L_H\sim  T^{-2\Delta_e/z},
$$
$$
\nu_N\sim B^{\beta-1}\mu^{\lambda-1}e^{\gamma+\delta} 
T^{-(2\beta+z(\lambda-1)+(1-\beta+\gamma+\delta)\Delta_e)/z}
$$
We mention, in passing, that the earlier 'no-go' theorem
regarding the possibility of constructing a sound one-parameter scaling 
theory of the cuprates \cite{chamon} was discussed under the overly narrow assumption 
of a particle-hole symmetric $d=3$ quantum-critical regime. 
Instead, in what follows we focus on the $d=2$ generic 
(including particle-hole asymmetric) case. 

Also, one should be alerted to the fact that the exponents governing the temperature dependencies in the kinetic coefficients
(8,9) would be the same as those appearing  
in their frequency-dependent optical counterparts.
However, the previous analysis based on the semiclassical kinetic equation warns that such leading (minimal) powers of $T$ may or may not actually survive, depending on whether or not 
the quasiparticle dispersion, Fermi surface topology, and spatial dimension 
conspire to provide for the comparable rates of the normal and 
umklapp inelastic scattering processes \cite{maslov}.  
It would appear, though, that in the cuprates, both, the multi-pocketed (in the under- and optimally-doped
cases) as well as the extended concave (in the over-doped case) hole Fermi surfaces should comply with the necessary conditions outlined in Ref.\cite{maslov}.

As an additional consistency check, 
Eqs.(9) agree with the Fermi liquid relations
\bea
S\sim {T\over e\sigma}{d\sigma\over d\mu},\nonumber\\
\nu_N\sim {T\over eB}{d\theta_H\over d\mu} 
\eea
which are both proportional to the Fermi surface curvature
but do not contain such single-particle characteristics as scattering time or effective mass
and, therefore, might be applicable beyond the coherent quasiparticle regime. 
Also, importantly, they hold for any values of the parameters
$\lambda, \beta, \gamma, \delta$
and independently of Eqs.(1-4). 

One can readily see that for
$[\Delta\rho/\rho]=2[\theta_H]$ the second and third of Eqs.(8) are only compatible, provided that 
\be
z\lambda+\Delta_e\gamma=0,
\ee
and then either of them can be used, alongside the two other relations, 
to determine the three unknowns: $z,\Delta_e$, and $\Delta_n$
\be
-2\beta+\Delta_e(\beta-\delta)=[\theta_H]=-2z,
\ee
$$
\Delta_n-2+2\Delta_e=[\sigma]=-z,
$$
$$
\Delta_n=[C]=z
$$
(any potential logarithmic factors would be missing here, though).

First, we discuss the solution of Eqs.(12)
with $\lambda=0$ which reads 
\be
z=\Delta_n=1-\Delta_e={\beta+\delta\over 2-\beta+\delta}
\ee
and satisfies (11) for arbitrary $\beta$ and $\delta$, if $\gamma=0$. For $\gamma\neq 0$, however, the solution can only exist, provided that 
$\beta=1$, and it turns out to be quite simple  
\be
z=1,~~\Delta_e=0,~~\Delta_n=1
\ee
These values are also physically relevant,
considering the natural linear field dependence
of the Hall angle, $\theta_H\sim {\sigma B/ en}$.

It can be readily seen
that the solution (14) yields the Seebeck, Hall Lorentz, and Nernst  coefficients (barring any Sondheimer-type cancellations) with the dimensions
\be
[S]=2z-1,~~[L_H]=2z-2,~~[\nu_N]=-1
\ee
These values obey the relation 
\be
[S]+[\nu_N]-[L_H]=0
\ee
which excludes the possibility of having $S\sim L_H$ 
(except for the unphysical $z\to\infty$ limit, see below).
In particular, for $\beta=1$ one obtains 
\be
S\sim T,~~L_H\sim const,~~\nu_N\sim T^{-1}
\ee
In the optimally doped cuprates, 
the experimentally measured thermopower, apart from a finite offset term, demonstrates a (negative) linear $T$-dependence, in agreement with (17) \cite{thermopower}.
As regards the Hall Lorentz and Nernst coefficients, the data on the untwinned samples of
the optimally doped $YBaCuO$ were fitted into a 
linear dependence, $L_H\sim T$, whereas $\nu_N$ was only 
found to decrease with increasing $T$, although no solid fit was 
provided \cite{ong}. 
Moreover, the Nernst signal increases dramatically with decreasing temperature, which effect 
has been attributed to the superconducting fluctuations and/or fluctuating vortex pairs whose (positive) contribution dominates over the  
quasiparticle one (which can be of either sign, depending on the dominant type of carriers) upon approaching $T_c$.
Besides, $\nu_N$ turns out to be strongly affected by a proximity to the 
pseudogap regime and can even become anisotropic \cite{nernst_pseudogap}.

It is worth mentioning, though, that the later Ref.\cite{matusiak} reported 
a slower temperature dependence of $L_H$ in the $LaSrCuO$, $EuBaCuO$, and $YBaCuO$ compounds. Specifically,
in the twinned $YBaCuO$ samples the dependencies 
$\sigma_H\sim T^{-2.7}$ and $\kappa_H\sim T^{-1.2}$ were found,
which is closer to the behavior predicted by Eq.(17).
While the origin of such experimental 
discrepancy remains unknown, it was also pointed out in Ref.\cite{matusiak} that their measurements of, both, $\sigma_H$ and $\alpha_H$ were 
carried out on the same sample, whereas the work of Ref.\cite{ong} was performed on the different ones. 

More experimental effort is clearly called for in order to ascertain 
the status of the above predictions. 
It should be mentioned, though, that the measurements of $L_H$ 
are technically quite a bit more involved than those of $\nu_N$. 

For $\lambda, \gamma\neq 0$ the consistency requirement (11) imposes a constraint on the coefficients
\be
\lambda(\beta+\delta)=2\gamma(\beta-1)
\ee
Provided that this condition is met, 
the solution of Eqs.(12) reads  
\be
z=\Delta_n=1-\Delta_e={\gamma\over \gamma-\lambda}, 
\ee
while the dimensions of the 
thermoelectric kinetic coefficients are still given by Eq.(15). 

For example, the solution with $\lambda=\gamma/2$ and $3\beta=\delta+4$
produces the exponents 
\be
z=2,~~\Delta_e=-1,~~\Delta_n=2
\ee
and the concomitant predictions 
\be
S\sim T^{3/2},~~L_H\sim T,~~\nu_N\sim T^{-1/2},
\ee 
which results match the linear $L_H$ (as well as, of course, Eqs.(1-4)). 
As to their practical relevance, 
the set $\lambda=-\beta=\gamma/2=\delta=-1$ would be the one suggested 
by the hydrodynamic relation $\theta_H\sim eBv^2\tau/\mu$ \cite{hydro}.

To rationalize the above solutions one might recall that 
in the context of the 
voluntarily generalized ('non-AdS/non-CFT') holographic approach the 
exponent $\Delta_n=d-\theta$ appears to be related
to the so-called 'hyperscaling-violation' parameter 
$\theta$ \cite{huijse,dvk2}. In the presence of a well-defined $d$-dimensional Fermi surface,
the latter is expected to coincide with its co-dimension, hence $\theta=d-1$,  
consistent with the solution (14). In contrast, the value $\theta=0$ implied by the solution (20) would hint at a point-like ('Dirac') Fermi surface (if any).

It is also interesting to compare the 'Fermi-surfaced' solution (14) with
the recent work of Ref.\cite{hk} which found 
\be
z=4/3,~~\Delta_e=-2/3,~~\Delta_n=2
\ee
at the expense of ignoring the implications of the particle-hole asymmetry.
Similar to (20), the solution (22) features $\theta=0$ and strong (infrared) 
charge renormalization ($\Delta_e<0$). However, the thermoelectric coefficients found in \cite{hk}
\be
S\sim T^{1/2},~~L_H\sim T,~~\nu_N\sim T^{-3/2}
\ee
do not obey Eqs.(15,16). Also, 
the agreement with the linear behavior of $L_H$
reported in Ref.\cite{ong} was guaranteed
by imposing it as one of the defining relations instead of Eq.(4).
In spite of this predestined success, however, this scheme  
fails to reproduce the linear (up to a constant) thermopower, although the authors claimed that the proposed $S=a-bT^{1/2}$ dependence 
could provide an even better fit to the data on $LaSrCuO$ \cite{thermopower}.  

Also, Eq.(22) predicts $C\sim T^{3/2}$
and, therefore, appears to be at odds 
with the observed thermodynamic properties as well \cite{specificheat}.
Moreover, it predicts the linear in $T$ 
(longitudinal) Lorentz ratio 
$
L={\kappa/T\sigma}
$
given by the same Eq.(23) as its Hall counterpart, contrary to 
a constant $L$, as suggested by the 
scenario (14). Experimentally, however, a slower-than-linear 
(electronic) Lorentz ratio has been reported \cite{minami}.

Also, in Ref.\cite{hk} the magnetic susceptibility was evaluated 
as the second field derivative of the free energy density   
\be
\chi_s={d^2f\over dB^2}\sim T^{(z+\Delta_n-2(2-\Delta_e))/z}
\ee
which 
yields the exponents $-2$ and $-3/2$ for the scenaria (14) and (22), 
respectively, thus providing additional 
means of discriminating between the two.

To that end, one can also invoke the charge susceptibility
\be
\chi_c={d^2f\over d\mu^2}\sim T^{(\Delta_n-z)/z}
\ee
which features the exponents $0$ and $1/2$ for the schemes (14) and (22), respectively, but, contrary to (24), conforms to 
the expectation of a constant Wilson ratio $C/\chi_cT$
in either case. 
 
Furthermore, in Ref.\cite{hk} the scheme (22) was argued to compare 
favorably with the energy- and momentum-dependent magnetic susceptibility
probed by inelastic neutron scattering.
Specifically, the limit   
\be
{\chi_s(\omega,{\bf q})\over \omega}|_{\omega\to 0}\sim |{\bf q}-{\bf q}_0|^{\Delta_n-2(2-\Delta_e)}
\ee
features the exponent $-10/3$ which was claimed to be close enough 
to the measured value of $-3$ \cite{mook}. 

One should, of course, be cautioned that the scaling (24) 
of the thermodynamic (i.e., uniform or ${\bf q}=0$) susceptibility 
does not necessarily characterize 
the spin response at the antiferromagnetic ordering vector ${\bf q}_0=(\pi,\pi)$.
It is interesting, though, that, such a caveat notwithstanding, the scheme
(14) yields the exact value of $-3$ which appears to be right on the data \cite{mook}.
Likewise, the momentum-integrated susceptibility 
\be
T\int d{\bf q}{\chi_s(\omega,{\bf q})\over \omega}|_{\omega\to 0}\sim T^{(\Delta_n+2\Delta_e+z-2)/z}
\ee
turns out to be constant in both cases of (14) and (22), again in agreement with the data \cite{mook}.

It is worth mentioning, though,
that, compared to the exotic exponents (22), those in Eq.(14) suggest a 
rather mundane physical picture where neither the Fermi liquid-like dispersion, nor 
the effective charge demonstrate any significant renormalization. 
Thus, constructing a viable phenomenological description of the optimally doped cuprates one might be able to do away without 
introducing the additional charge exponent $\Delta_e$,  
contrary to the assertions made in Refs.\cite{karch,hk,guteraux}. 

Besides, conspicuously enough, the value of $z$ from Eq.(22) 
has not appeared in any of the (ostensibly) pertinent 
literature \cite{gauge}. 

The general scaling scheme (8,9) can be further adapted to explore the 
alternate scenaria in which the extended Fermi surface  
is replaced by nodal (Dirac-like) points in, both, 
superconducting and pseudogap phases. 
Such models have been invoked in the analyses of the underdoped cuprates where non-analytical magnetic field dependences are expected, such as 
$
\kappa_H\sim B^{1/2}T
$ 
observed in Ref.\cite{ong}, except at the lowest fields. 

In this case, one would be prompted to use the value 
$\beta=1/2$ in Eqs.(8,9) which is characteristic
of the field-induced density of the nodal excitations \cite{simon}. 
Also, any scaling scheme adequate to the underdoped regime
would have to incorporate the inelastic 
quasiparticle scattering rate $\sim T^2$, 
as revealed by the recent experiments \cite{underdoped}.

The scaling equations (8,9) can also be used 
to fit the power-law optical kinetic coefficients.
To that end, it would be interesting
to investigate the possibility of matching 
the anomalous power-law decay of the  
high-frequency optical conductivity, $\sigma\sim\omega^{-2/3}$, \cite{optical} (which the holographic work of Ref. \cite{horowitz}
claimed to have reproduced numerically, albeit only 
over less than half a decade) 
as well as its compatibility with the conjectured sum rule for its Hall 
counterpart \cite{drew}.

Returning to the recent holographic theories 
of the cuprates, one should mention the well-known proposal
$
\sigma\sim 1/{\cal S}
$
where $\cal S$ is the entropy density \cite{zaanen}.
The targeted dependence (1) would then arise if $\cal S$ 
(which scales in the same way as the specific heat - or, for that matter,
quasiparticle density) were linear in $T$, again consistent with (14). 

One should be reminded, though, that the above dependence was 
actually obtained from the expression 
\be
\sigma\sim T^{(\theta-2-d)/z}\sim {1\over {\cal S}}T^{-2/z}
\ee
where the desired behavior (1) 
sets in by taking the limit $z\to\infty$ while keeping the entropy factor intact. Such limit has been 
abundantly discussed in the holographic literature 
and argued to describe various 'locally-critical' systems. 
   
More specifically, Eq.(28) was shown to emerge in the strong coupling regime, whereas the perturbative result is non-universal, 
$
\sigma\sim T^{2(z-1-\Delta_O)/z}
$, 
and determined by the dimension of the operator $O$ 
breaking momentum conservation. It restores Eq.(28) only 
when the Harris criterion
of the relevance of disorder, $\Delta_O=(d-\theta)/2+z$, is met \cite{sachdev}. 

As an alternative to the 'locally critical' regime,   
another potential candidate to the holographic 
theory of the cuprates, the hyperscaling violation model with $z=3/2$ and $\theta=1$, has been proposed \cite{huijse,dvk2}. 
However, without any sleight of hands
the conductivity exponent 
($-3/z=-2$) given by Eq.(28) would be off the target, just as well.

Thus, currently, a reliable calculation of the
conductivity (as well as the other kinetic coefficients) 
within the same holographic model  
allowing for a systematic comparison with the data  
is still awaiting to be performed.

To summarize, in this work   
we proposed a systematic scaling approach for constructing 
viable phenomenologies of the normal state of the cuprates by making a proper account
of the particle-hole asymmetry. In particular, we focused on two plausible solutions given 
by Eqs.(14) and (20) which account for a variety of the algebraic temperature dependencies that have already been established experimentally. As regards those 
observables whose scaling behaviors have not yet been reliably determined,
including the Hall Lorentz and Nernst coefficients, we make a concrete 
prediction (16) for the relation between their critical exponents and that of the thermopower.   

Also, our predictions are contrasted against the alternate 
scheme of Ref.\cite{hk} inspired by the holographic theories which, unlike our approach, 
does not properly account for the particle-hole asymmetry and, therefore, 
violates the relation (16). Also, apart from the questionable behavior of the specific heat, thermopower, and spin susceptibility, it predicts a rather exotic value of the dynamical  
critical exponent and a strong temperature dependence of 
the effective charge given by Eq.(22).

Such concrete predictions may facilitate a systematic comparison 
between the competing approaches, as well as experiment.   
To that end, the host of available scaling data on the cuprates 
provides a means of ascertaining the true status of the intriguing, yet speculative holographic ideas which have been proclaimed as a novel powerful technique for handling the general problem of strong electron correlations in the cuprates and other 'strange metals'. 

The author acknowledges the Aspen Center for Physics
funded by the grant NSF under Grant 1066293 and the Galileo Galilei Institute (Florence, Italy)
for their hospitality and workshop participation support.

\end{document}